\documentclass[11pt]{article}

\usepackage{amsmath, amssymb, mathrsfs, bm, latexsym,graphics,color, graphicx,url,floatflt}

\newcommand{\ignore}[1]{}

\newtheorem{theorem}{Theorem}[section]

\newtheorem{proposition}[theorem]{Proposition}
\newcommand{\blackslug}{\penalty 1000\hbox{
    \vrule height 8pt width .4pt\hskip -.4pt
    \vbox{\hrule width 8pt height .4pt\vskip -.4pt
          \vskip 8pt
      \vskip -.4pt\hrule width 8pt height .4pt}
    \hskip -3.9pt
    \vrule height 8pt width .4pt}}

\newenvironment{proof}{$\;$\newline \noindent {\sc Proof.}$\;\;\;$\rm}{\qed}
\newcommand{\qed}{\hspace*{\fill}\blackslug}

\def\boxit#1{\vbox{\hrule\hbox{\vrule\kern4pt
  \vbox{\kern1pt#1\kern1pt}
\kern2pt\vrule}\hrule}}

\begin{document}

\title{An exact algorithm with the time complexity of $O^*(1.299^m)$  for the weighed mutually exclusive set cover problem}

\author{Songjian Lu and Xinghua Lu}

\date{}

\maketitle

\vspace{-0.8cm}
\begin{center} Department of Biomedical Informatics,

University of Pittsburgh, Pittsburgh, PA 15219, USA

Email: songjian@pitt.edu, xinghua@pitt.edu
\end{center}

\begin{abstract}
In this paper, we will introduce an exact algorithm with a time
complexity of $O^*(1.299^m)^{\dag}$
\let\thefootnote\relax\footnotetext{$^{\dag}${\bf Note:} Following
the recent convention, we use a star $*$ to represent that the
polynomial part of the time complexity is neglected.} for the {\sc
weighted mutually exclusive set cover} problem, where $m$ is the
number of subsets in the problem. This problem has important
applications in recognizing mutation genes that cause different
cancer diseases.
\end{abstract}

\section{Introduction}
The {\sc set cover} problem is that: given a ground set $X$ of $n$
elements and a collection ${\cal F}$ of $m$ subsets of $X$, try to
find a minimum number of subsets $S_1,S_2,\ldots,S_h$ in ${\cal
F}$ such that $\cup_{i=1}^hS_i = X$. If we add an additional
constrain such that all subsets in the solution are pairwise
disjoint, then the {\sc set cover} problem becomes the {\sc
mutually exclusive set cover} problem. If we further assign each
subset in ${\cal F}$ a real number weight and search the solution
with the minimum weight, i.e. the sum of weights of subsets in the
solution is minimized, then the problem becomes the {\sc weighted
mutually exclusive set cover} problem.

Recently, the {\sc weighted mutually exclusive set cover} problem
has found important applications in cancer study to identify
driver mutations~\cite{Ciriello,Miller}, i.e. somatic mutations
that cause cancers. As somatic mutations will change the
structures (and therefore the functions) of signaling proteins;
thus, perturb cancer pathways that regulate the expressions of
genes in certain important biological processes, such as cell
death, cell proliferation etc. The perturbations within a common
cancer pathway are often found to be mutually exclusive in a
single cancer cell, i.e. each tumor usually has only one
perturbation on one given cancer pathways (one perturbation is
enough to cause the disease; hence, there is no need to wait for
another perturbation). Modern lab techniques can identify somatic
mutations and gene expressions of cancer cells. After
preprocessing the data, we will obtain following information for
important biological processes, e.g. cell death: 1)which cancer
cells have disturbed the expressions of genes in the biological
process; 2) which genes have been mutated in those cancer cells;
3) how possible each mutation is related to the given biological
process (i.e. each mutation is assigned a real number weight).
Then next step is finding a set of mutations such that each cancer
cell has one and only one mutation in the solution set (mutually
exclusive) and the sum of weights of all genes in the solution set
is minimized, which is the {\sc weighted mutually exclusive set
cover} problem.

While there is not much research on the {\sc mutually exclusive
set cover} or the {\sc weighted mutually exclusive set cover}
problems, the {\sc set cover} problem has been paid much
attention. The {\sc set cover}, which is equivalent to the  {\sc
hitting set} problem, is a fundamental NP-hard problem in Karp's
21 NP-complete problems~\cite{Karp1972}. One research direction
for the {\sc set cover} problem is approximation algorithms, e.g.
papers ~\cite{alon,feige,Kolliopoulos,lund} gave polynomial time
approximation algorithms that find solutions whose sizes are at
most $c\log n$ times the size of the optimal solution, where $c$
is a constant. Second direction is using $k$, the number of
subsets in the solution, as parameter to design fixed-parameter
tractable (FPT) algorithms for the equivalent problem, the {\sc
hitting set} problem. Those algorithms have a constrain such that
each element in $X$ is included in at most $d$ subsets in ${\cal
F}$, i.e. sizes of all subsets in the {\sc hittng set} problem are
upper bound by $d$; it is also called the {\sc $d$-hitting set}
problem. For example, paper~\cite{niedemedier} gave an
$O^*(2.270^k)$ algorithm for the {\sc $3$-hitting set} problem,
and paper~\cite{fernau_2} further improved the time complexity to
$O^*(2.179^k)$. The third direction is designing algorithms that
use $n$ as parameter in the condition that $n$ is much less than
$m$. Papers~\cite{bjorklund,Hua2} designed algorithms with time
complexities of $O^*(2^n)$ for the problem. The
paper~\cite{bjorklund} also extended the algorithm to solve the
{\sc weighted mutually exclusive set cover} problem with the same
time complexity. Paper \cite{Lu2011} improved the time complexity
to $O^*(2^{\frac{\log_2d}{1+\log_2d}n})$ under the condition that
at least $\frac{n}{1+\log_2n}$ elements in $X$ are included in at
most $d$ subsets in ${\cal F}$. This algorithm can also be
extended to the {\sc weighted mutually exclusive set cover}
problem with the same time complexity. However, in the application
of cancer study, neither $n$ is less than $m$ nor each element in
$x$ is included in bounded number of subsets in ${\cal F}$. Hence,
there is a need to design new algorithms.

In this paper, we will design a new algorithm that uses $m$ as
parameter (in application of cancer study, $m$ is smaller than
$n$, where $n$ can be as large as several hundreds). Trivially, if
using $m$ as parameter, we can solve the problem in time of
$O^*(2^m)$, where the algorithm basically just tests every
combination of subsets in ${\cal F}$. To our best knowledge, we
have not found any algorithm that is better than the trivial
algorithms when using $m$ as parameter. This paper will give the
first un-trivial algorithm with the time complexity of
$O^*(1.299^m)$ to solve the {\sc weighted mutually exclusive set
cover} problem. We have tested this algorithm in the cancer study,
and the program can finish the computation practically when $m$ is
less than 100.

\section{The {\sc weighted mutually exclusive set cover} problem is NP-hard}
The formal definition of the {\sc weighted mutually exclusive set
cover} problem is: given a ground set $X$ of $n$ elements, a
collection ${\cal F}$ of $m$ subsets of $X$, and a weight function
$w: {\cal F} \rightarrow [0, \infty)$, if ${\cal F'}
=\{S_1,S_2,\ldots,S_h\} \subset {\cal F}$ such that
$\cup_{i=1}^hS_i=X$, and $S_i\cap S_j=\emptyset$ for any $i \neq
j$, then we say ${\cal F'}$ is a mutually exclusive set cover of
$X$ and $\sum_{i=1}^hw(S_i)$ is the weight of ${\cal F'}$; the
goal of the problem is to find a mutually exclusive set cover of
$X$ with the minimum weight, or report that no such solution
exists.

As we have not found the proof of NP-hardness for the {\sc
weighted mutually exclusive set cover} problem, in this section,
we will prove that the {\sc mutually exclusive set cover} problem
is NP-hard; thus, prove that the {\sc weighted mutually exclusive
set cover} problem is NP-hard.

We will prove the NP-hardness of the {\sc mutually exclusive set
cover} problem by reducing another NP-hard problem, the {\sc
maximum set packing} problem, to it. Remember that the {\sc
maximum set packing} problem is: given a collection ${\cal F}$ of
subsets, try to find an ${\cal S} \subset {\cal F}$ such that
subsets in ${\cal S}$ are pairwise disjoint and $|{\cal S}|$ is
maximized.

\begin{theorem}
The {\sc mutually exclusive set cover} problem is NP-hard.
\begin{proof}
Let ${\cal S} = \{S_1,S_2,\ldots,S_m\}$ be an instance of the {\sc
maximum set packing} problem, where $X'=\cup_{i=1}^mS_i =
\{x_1,x_2,\ldots,x_n\}$. We create an instance of the {\sc mutually
exclusive set cover} problem such that:
\begin{itemize}
\item $X = X' \cup \{T_1,T_2,\ldots,T_m\}$, where $T_i = \{t_{i1},
t_{i2},\ldots,t_{i(n+1)}\}$ for all $1 \leq i \leq m$;

\item ${\cal F} = {\cal F'} \cup {\cal F''} \cup {\cal F'''}$,
where ${\cal F'}=\{\{x_1\},\{x_2\},\ldots,\{x_n\}\}$, ${\cal F''}=
\{S_1\cup T_1,S_2\cup T_2, \ldots, S_m\cup T_m\}$, and ${\cal
F'''} =
\cup_{i=1}^m\{\{t_{i1}\},\{t_{i2}\},\ldots,\{t_{i(n+1)}\}\}$.
\end{itemize}
Next, we will prove that if ${\cal P} = \{P_1,P_2,\ldots, P_k\}$
is a solution of the {\sc mutually exclusive set cover} problem,
then ${\cal S'} = \{S'_1,S'_2,\ldots, S'_{k'}\}$ is a solution of
the {\sc maximum set packing} problem, where ${\cal P}\cap {\cal
F''} = \{S'_1\cup T'_1, S'_2\cup T'_2, \ldots, S'_{k'}\cup
T'_{k'}\}$. Thus we will prove that the time to solve the  {\sc
maximum set packing} problem is bounded by the total time of
transforming the {\sc maximum set packing} problem into the {\sc
mutually exclusive set cover}, and of solving the {\sc mutually
exclusive set cover} problem. Therefore, the {\sc mutually
exclusive set cover} problem is NP-hard.

As subsets in ${\cal P}$ are pairwise disjoint, it is obvious that
subsets in ${\cal S'}$ are pairwise disjoint. Hence, if we suppose
that ${\cal S'}$ is not the solution of the {\sc maximum set
packing} problem, then there must exists a ${\cal S''} =
\{S''_1,S''_2, \ldots, S''_{k'}\} \subset {\cal S}$ such that
subsets in ${\cal S''}$ are pairwise disjoint and $k'>k$. Thus we
can make a new solution ${\cal P'}$ of the {\sc mutually exclusive
set cover} problem such that ${\cal P'}$ includes $\{S''_1\cup
T''_1, S''_2\cup T''_2, \ldots, S''_{k'}\cup T''_{k'}\} \subset
{\cal F''}$ and other subsets in ${\cal F'}$ and ${\cal F'''}$. If
let $|X' - \cup_{i=1}^kS'_i| = n_1$ and $|X' -
\cup_{i=1}^{k'}S''_i| = n_2$ (Note: any $T_i$, which is not
covered by a subset in ${\cal F''}$, needs $n+1$ subsets in ${\cal
F'''}$ to cover it; any $x_i \in X'$, which is not covered by a
subset in ${\cal F''}$, needs a subset in ${\cal F'}$ to cover
it), then
\[ |{\cal P}| = k + (m-k)(n+1) + n_1,~~ \]
and
\[|{\cal P'}| = k' + (m-k')(n+1) + n_2. \]
Therefore $|{\cal P}|-|{\cal P'}| = (k'-k)n+n_1-n_2>0$, i.e.
${\cal P'}$ is a solution with less subsets in ${\cal F}$, which
cases contradiction that ${\cal P}$ is the solution of the {\sc
mutually exclusive set cover} problem. Hence, ${\cal S'}$ is a
solution of the {\sc maximum set packing} problem.
\end{proof}\end{theorem}

\section{The main Algorithm}

In this section, we will introduce our new algorithm to solve the
{\sc weighted mutually exclusive set cover} problem.

\begin{figure}[ht]
\centering \scalebox{0.8}{\includegraphics{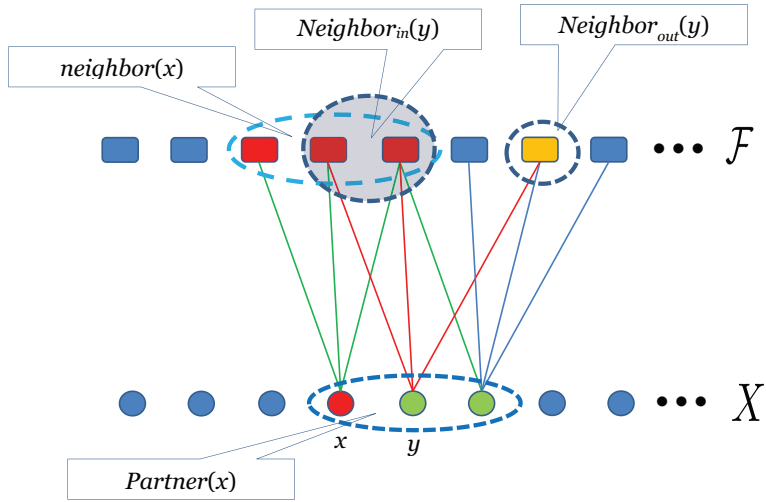}}
\caption{Graph representation and some notations of the problem}
\label{fig_1}
\end{figure}

 Let
$(X,{\cal F},w)$ be an instance of the {\sc weighted mutually
exclusive set cover} problem. We can use a bipartite graph to
represent $(X,{\cal F},w)$ such that all nodes on one sides are
subsets in ${\cal F}$ while nodes on the other side are elements
in $X$, and if an element $u$ of $X$ is in subset $U$, i.e. $u \in
U$, then an edge is added between $u$ and $U$. For the
convenience, let us introduce some notations. The
Figure~\ref{fig_1} can help you to understand and remember
following notations.

For any $x \in X$, let $neighbor(x) = \{S| S \in {\cal F} \text{
and } x \in S\}$, $degree(x)=|neighbor(x)|$, $partner(x) = \cup_{S
\in neighbor(x)}S$. For any $y$ in $partner(x)$, let
$neighbor_{in} = neighbor(y) \cap neighbor(x)$, $degree_{in}(y) =
|neighbor_{in}(y)|$, $neighbor_{out} = neighbor(y)-neighbor(x)$,
$degree_{out}(y) = |neighbor_{out}(y)|$.

\begin{figure}[htb]
\footnotesize
\begin{tabbing}
xxxx\=xx\=xx\=xx\=xx\=xx\=xx\=xx\=xx\=xx\=xx\=xx\=xx\=\kill
{\bf Algorithm-1 WMES-Cover$((X,{\cal F},w), Solution_{partial}, Solution_{final}))$}\\
\textbf{Input:} An instance of the {\sc weighted mutually exclusive set cover} problem, two variables,\\
\>\>\> where $Solution_{final}$ is a global variable to keep the
best solution.\\
\textbf{Output:} A minimum weight mutually exclusive set cover or ``No Solution".\\
\\

1 \> {\bf if} $X==\emptyset$ {\bf then}\\
1.1 \>\> {\bf if} $weight(Solution_{partial})<weight(Solution_{final})$ {\bf then} replace $Solution_{final}$ with $Solution_{patial}$;\\
2 \>Find $x \in X$ such that $d=degree(x)$ is minimized;\\
3 \>{\bf if} $d==0$ {\bf then} {\bf return} ``No Solution";\\
4 \>{\bf if} $d==1$ {\bf then} WMES-Cover$((X-\{x\},{\cal
F}-neighbor(x),w),Solution_{partial}\cup neighnor(x),
Solution_{final})$;\\
5 \> {\bf if} $degree_{out}(y)==0$ for all $y \in partner(x)$ {\bf
then}\\
5.1 \>\> {\bf if} there exists $S \in neighbor(x)$ such that
$S==partner(x)$ {\bf then}\\
5.1.1 \>\>\> WMES-Cover$((X-S,{\cal
F}-neighbor(x),w),Solution_{partial}\cup
\{S\}, Solution_{final})$;\\
 \>\> {\bf else}\\
5.1.2 \>\>\> {\bf return} ``No Solution";\\
6 \> {\bf if} $d==2$ {\bf then} // Suppose $neighbor(x)=\{S_1,S_2\}$; note that $S_1\subset X$ and $S_2\subset X$.\\
6.1 \>\> WMES-Cover$((X-S_1,{\cal F}-\cup_{u \in
S_1}neighbor(u),w),Solution_{partial}\cup
\{S_1\}, Solution_{final})$;\\
6.2 \>\> WMES-Cover$((X-S_2,{\cal F}-\cup_{u \in
S_2}neighbor(u),w),Solution_{partial}\cup
\{S_2\}, Solution_{final})$;\\
 \> {\bf else} // (Note: $d>2$)\\
6.3 \>\> {\bf if} there exists a $y \in
partner(x)$ such that $degree_{out}(y) =  1$ {\bf then}\\
6.3.1 \>\>\> Let $y\in partner(x)$ such that $degree_{out}(y)=1$ and $W'\in neighbor_{out}(y)$; \\
6.3.2 \>\>\> {\bf if} $| neighbor(x)-neighbor(y)|>0$ {\bf then} //
(Note: $|
neighbor(x)-neighbor(y)|\leq 1$) \\
6.3.2.1 \>\>\>\> Find any $W \in neighbor(x)-neighbor(y)$;\\
6.3.2.2 \>\>\>\> WMES-Cover$((X-W'\cup W,{\cal F}-\cup_{u \in
W'\cup W}neighbor(u),w),Solution_{partial}\cup
\{W',W\}, Solution_{final})$;\\
6.3.2.3 \>\>\>\> WMES-Cover$((X,{\cal F}-\{W',W\},w),Solution_{partial}, Solution_{final})$;\\
\>\>\> {\bf else}\\
6.3.2.4 \>\>\>\> Find any $W \in neighbor(x)$;\\
6.3.2.5 \>\>\>\> WMES-Cover$((X-W,{\cal F}-\cup_{u \in
W}neighbor(u),w),Solution_{partial}\cup
\{W\}, Solution_{final})$;\\
6.3.2.6 \>\>\>\> WMES-Cover$((X,{\cal F}-\{W',W\},w),Solution_{partial}, Solution_{final})$;\\

\>\> {\bf else}\\

6.3.3 \>\>\> Find a $y \in
partner(x)$) such that $degree_{out}(y)$ is maximized;\\
6.3.4 \>\>\> Find a $Z\in neighbor_{in}(y)$; \\
6.3.5 \>\>\> WMES-Cover$((X-Z,{\cal F}-\cup_{u \in
Z}neighbor(u),w),Solution_{partial}\cup
\{Z\}, Solution_{final})$;\\
6.3.6 \>\>\> WMES-Cover$((X,{\cal F}-\{Z\},w),Solution_{partial}, Solution_{final})$;\\
\end{tabbing}
\vspace*{-3mm}
\caption{Algorithm for the {\sc weighted mutually exclusive set
cover} problem.} \label{Algorithm_main}
\end{figure}

The main algorithm, Algorithm-1, is shown in
Figure~\ref{Algorithm_main}. Basically, the Algorithm-1 first
finds an $x \in X$ with minimum degree and then branches at one
subset in $neighbor(x)$ (such as in step 6.2.2 and 6.2.3). For the
convenience, if $degree(x)=d$, then we say that Algorithm-1 is
doing a $d$-branch. Because of steps 3,4,5, when the program
arrives at step 6, we must have: 1) $d=degree(x)\geq 2$; 2) for
any $u \in X$, $degree(u) \geq d$; 3) there exists a $y\in
partner(x)$ such that $degree_{out}(y)>0$.

The Algorithm-1 is basically searching the solution by going
through a search tree; hence, if knowing the number of leaves in
the search tree, then we will obtain the time complexity of the
Algorithm-1. Next, we will estimate the number of leaves in the
search tree by studying the different cases of branching. We begin
from the $2$-branch.

\begin{proposition}\label{main_PR_2} The search tree has at most $1.273^m$ leaves If
only the 2-branches are applied in Algorithm-1.
\begin{proof}
Suppose that $degree(x)=2$ and $y \in partner(x)$ such that
$degree_{out}(y)>0$. Let $neighbor(x) = \{S_1,S_2\}$.

In the case of $degree_{out}(y)=1$, let $neighbor_{out}(y) =
\{S''\}$. In the branches of choosing either $S_1$ or $S_2$ into
the solution, if $y$ is covered, then $S''$ will be removed from
the ${\cal F}$, or else if $y$ is not covered yet, then $S''$ will
be chosen into the solution in order to cover $y$ (note: after
$S_1,S_2$ are removed, $degree(y)=1$ in the new instance (at line
6.1.1 and 6.1.2 of Algorithm-1); thus, $S''$ will be included into
the solution in the next call of the Algorithm-1 in this branch).
Hence, in any case, $3$ subsets in ${\cal F}$ will be removed. If
letting $T(k)$ be the number of leaves in the search tree when
$|{\cal F}|=k$, then we will obtain the following recurrence
relation
\[T(k) \leq 2T(k-3).~~~~~~~~~~~~~~~~~~~~~~~~~~~~~~~~~~~~~~~~~~~~~~~~~\text{(1)}\]
The characteristic equationof this recurrence relation is
$r^3-2=0$ $^{\ddag}$
\let\thefootnote\relax\footnotetext{$^{\ddag}${\bf Note:} Given a
recurrence relation $T(k) \leq \sum_{i=0}^{k-1}c_iT(i)$ such that
all $c_i$ are nonnegative real numbers, $\sum_{i=0}^{k-1}c_i>0$,
and $T(0)$ represents the leaves, then $T(k) \leq r^k$, where $r$
is the unique positive root of the characteristic equation $t^k -
\sum_{i=0}^{k-1}c_it^i=0$ deduced from the recurrence
relation~\cite{chen}.}; hence, we will have $T(m)<1.260^m$.

In the case of $degree_{out}(y)>1$, we consider following
sub-cases.

{\it Sub-case 1}. Suppose $degree_{in}(y)=1$, and $y \in S_1$.
Then at least $S_1$ and $S_2$ will be removed from ${\cal F}$ for
the branch of choosing $S_2$ into the solution; at least $S_1$,
$S_2$, and all subsets (at least two) in $neighbor_{out}(y)$ will
be removed for the branch of choosing $S_1$ into the solution.
Thus the recurrence relation of $T(k)$ is
\[ T(k) \leq T(k-2) + T(k-4).~~~~~~~~~~~~~~~~~~~~~~~~~~~~~~~~~~~~~\text{(2)}  \]which leads to $T(m)<1.273^m$.

{\it Sub-case 2}. Suppose $degree_{in}(y)=2$. Then in either
branch, $y$ is covered by $S_1$ or $S_2$, which is chosen into the
solution. Hence, $S_1,S_2$, and all subsets (at least two) in
$neighbor_{out}(y)$ will be removed from ${\cal F}$. Thus we will
obtain the recurrence relation
\[ T(k) \leq 2T(k-4).~~~~~~~~~~~~~~~~~~~~~~~~~~~~~~~~~~~~~~~~~~~~~~~~~~~~\text{(3)} \]which leads to $T(m)<1.190^m$.

By considering all above cases, we obtain that $T(m)\leq 1.273^m$.
\end{proof}\end{proposition}

Now, we consider the case of doing $3$-branch. Remember that when
Algorithm-1 is doing a $3$-branch, $degree(x) \geq 3$ for all $x
\in X$.

\begin{proposition}\label{main_PR_3} The search tree has at most $1.299^m$ leaves If
only the $d$-branches for $d<=3$ are applied in Algorithm-1.
\begin{proof}
The cases of $2$-branches are considered in the last proposition.
Now we consider the cases of $3$-branches. Suppose that
$degree(x)=3$ and $y \in partner(x)$ such that
$degree_{out}(y)>0$. Let $neighbor(x)=\{S_1,S_2,S_3\}$.

If $degree_{out}(y)=1$, then $degree_{in}(y) \geq 2$ (as
$degree(y) \geq 3$). Let $\{S'\}=neighbor_{out}(y)$. We further
consider following sub-cases.

{\it Sub-case 1}. Suppose $degree_{in}(y)=2$. Let $S_1 \in
neighbor(x)-neighbor(y)$. The Algorithm-1 branches at $S_1$. The
branch one includes $S_1$ into the solution; thus, $S_2,S_3$ will
be removed. This will further make $degree(y) = 1$. Hence, $S'$
will also be included into the solution. Totally, in this branch,
we will remove at least $4$ subsets from ${\cal F}$. In branch
two, we will exclude $S_1$ from the solution. Then either $S_2$ or
$S_3$ must be included into the solution. Thus $y$ is covered by
$S_2$ or $S_3$, and $S'$ will not be in the solution. Therefore,
in this branch, we know that at least $S_1$ and $S'$ will be
removed. So we will obtain the recurrence relation
\[ T(k) \leq T(k-2) + T(k-4),~~~~~~~~~~~~~~~~~~~~~~~~~~~~~~~~~~~~~\text{(4)}  \]
which leads to $T(m)<1.273^m$.

{\it Sub-case 2}. Suppose $degree_{in}(y)=3$. Then $S'$ will not
in the solution and any one of $S_1,S_3,S_3$ (one and only one of
them must be included into the solution to cover $x$) will cover
$y$. The Algorithm-1 will branch at any one of $S_1,S_2,S_3$.
Without loss of generality, we branch at $S_1$. In the branch of
including $S_1$ into the solution, $S_1,S_2, S_3$ will be removed,
which will totally remove at least $4$ subsets. In the branch of
excluding $S_1$ into the solution, $S_1$ will be removed. Thus $2$
subsets will be removed. We will obtain the following recurrence
relation
\[ T(k) \leq T(k-2) + T(k-4),~~~~~~~~~~~~~~~~~~~~~~~~~~~~~~~~~~~~~\text{(5)}  \]
which leads to $T(m)<1.273^m$.

In the case of $degree_{out}(y)>1$, Let $S_1 \in
neighbor_{in}(y)$. Algorithm-1 branches at $S_1$. In the first
branch, $S_1$ is included into the solution. Then $S_1,S_2,S_3$
and at least $2$ subsets in $neighbor_{out}(y)$ will be removed.
In the second branch, $S_1$ is excluded, which will make
$degree(x)=2$ in the new instance; hence, in this branch, a
$2$-branch will follow. Thus even considering the worst case of
the $2$-branch (the recurrence relation (2)), we will have
\[T(k)\leq 2T(k-5)+ T(k-3), ~~~~~~~~~~~~~~~~~~~~~~~~~~~~~~~~~~~~~~~\text{(6)}\]
which will lead to $T(m) \leq 1.299^m$.

From all above cases and Proposition~\ref{main_PR_2}, we will have
$T(m) \leq 1.299^m$.
\end{proof}\end{proposition}

Let us consider the case of doing $d$-branch for $d>3$.

\begin{proposition}\label{main_PR_4} The search tree in Algorithm-1 has at most $1.299^m$ leaves.
\begin{proof}
We only need to consider the cases of $d$-branches for $d>3$.
Suppose that $degree(x)=d$ and $y \in partner(x)$ such that
$degree_{out}(y)>0$. Let $neighbor(x)=\{S_1,S_2,\ldots,S_d\}$.

In the case of $degree_{out}(y)=1$, $degree_{in}(y)$ can only be
$d-1$ or $d$.

{\it Sub-case 1}. Suppose $degree_{in}(y)=d-1$. Then there is one
and only one subset in $neighbor(x)-neighbor_{in}(y)$. Without
loss of generality, we suppose $S_1 \not\in neighbor_{in}(y)$.
Algorithm-1 will branch on $S_1$ such that in the branch of
including $S_1$ into the solution, all $d$ subsets in
$neighbor(x)$ and one subset in $neighbor_{out}(y)$ will be
removed (i.e. in this branch, at least $5$ subsets will be
removed; in the branch of excluding $S_1$ from the solution, one
subset in $\{S_2,S_3,\ldots,S_d\}$ will be included into the
solution, which $y$ will be covered and the only subset in
$neighbor_{out}(y)$ will be removed (i.e. in this branch, two
subsets will be removed). Therefore, we will have following
recurrence relation
\[ T(k) \leq T(k-5) + T(k-2),~~~~~~~~~~~~~~~~~~~~~~~~~~~~~~~~~~~~~\text{(7)}  \]
which leads to $T(m)<1.237^m$.

{\it Sub-case 2}. Suppose $degree_{in}(y)=d$.  Without loss of
generality, we suppose that Algorithm-1 branches on $S_1$. Then it
is easy to understand the we will have the following recurrence
relation
\[ T(k) \leq T(k-5) + T(k-2),~~~~~~~~~~~~~~~~~~~~~~~~~~~~~~~~~~~~~\text{(8)}  \]
which leads to $T(m)<1.237^m$.

In the case of $degree_{out}(y)>1$, suppose $S_1 \in
degree_{in}(y)$ and Algorithm-1 branches on $S_1$. Then in the
branch of including $S_1$ into the solution, all subsets in
$neighbor(x)$ and $neighbor_{out}(y)$ will be removed (at least
$6$ subsets will be removed). In the branch of excluding $S_1$
into the solution, at least one subset $S_1$ will be removed.
Hence, we will have the recurrence relation
\[ T(k) \leq T(k-6) + T(k-1),~~~~~~~~~~~~~~~~~~~~~~~~~~~~~~~~~~~~~\text{(9)}  \]
which leads to $T(m)<1.286^m$.

Considering all above cases, Proposition~\ref{main_PR_2}, and
Proposition~\ref{main_PR_3}, we have $T(m)\leq 1.299^m$.
\end{proof}\end{proposition}

\begin{theorem}\label{main_TH_2}
The {\sc weighted mutually exclusive set cover} problem can be
solved by an algorithm with a time complexity of $O^*(1.299^m)$.
\begin{proof}
Let $({\cal F}, X, w)$ be an instance of the {\sc weighted
mutually exclusive set cover} problem, where $X$ is a ground set
of $n$ elements, $\cal{F}$ is a collection of $m$ subsets of $X$,
and $w: {\cal F} \rightarrow [0, \infty)$ is the weight function.
Now we prove that the problem can be solved by the Algorithm-1 in
time $O^*(1.299^m)$.

The correctness of the algorithm is easy to understand. If there
is an $x \in X$ such that $degree(x)=0$, then $x$ cannot be
covered by any subset in ${\cal F}$. Thus, the problem has no
solution. The step 3 of the Algorithm-1 deals with this situation.
If, for any given $x\in X$, $degree(x)=1$, then there exists one
and only one subset in ${\cal F}$ that covers $x$, i.e.
$neighbor(x)$ must be included into the solution. Thus $x$ and
$neighbor(x)$ will be removed from the problem. This situation is
dealt with in step 4. If for all $y$ in $partner(x)$,
$degree_{out}(y)=0$, then $partner(x)$ can only be covered by
subset(s) in $neighbor(x)$. By the exclusivity, at most one subset
in $neighbor(x)$ can be chosen into the solution. Thus, if finding
a subset $S$ in $neighbor(x)$ such that $S=partner(x)$, then
Algoirhtm-1 will include $S$ into the solution, or else the
problem has no solution. The step 5 of the Algorithm-1 deals with
this situation.

After the Algorithm-1 reaches step 6, we have: 1) for all $x' \in
X$, $degree(x')\geq degree(x)>1$ (as $x$ is the element in $X$
with the minimum degree); 2) there is a $y\in partner(x)$ such
that $degree_{out}(y)>0$. If $d=neighbor(x)=2$, then one and only
one subset in $neighbor(x)$ will be in the solution. The step 6.1
and 6.2 correctly deals with this situation. For the cases after
step 6.2, the Algorithm-1 basically chooses one subset $S$ in
$neighbor(x)$ and branches on $S$ such that one branch includes
$S$ into the solution and the other branch excludes $S$ from the
solution (Note: when $degree_{out}(y)=1$, we used a small trick to
include or exclude the additional subset in $neighbor_{our}(y)$
into or from the solution; please refer to sub-case 1 and sub-case
2 in the Proposition~\ref{main_PR_4}). Therefore, Algorithm-1 will
go through the search tree and find the solution with the minimum
weight (if the solution exists), which is saved in step 1.1.

By Proposition~\ref{main_PR_4}, the search tree has at most
$1.299^m$ leaves. Hence, the time complexity of the algorithm is
bounded by $O^*(1.299^m)$. If we further notice that the time to
process each node is bounded by $O(mn)$, then the more accurate
time complexity of the algorithm is $O(1.299^mmn)$.
\end{proof}\end{theorem}

\section{Problem extension}
In this paper, we first proved that the {\sc weighted mutually
exclusive set cover} problem is NP-hard. Then we designed the
first non-trivial algorithm, which uses the $m$ as parameter, with
a time complexity of $O^*(1.352^m)$ for the problem. the {\sc
weighted mutually exclusive set cover} problem has been used to
find the driver mutations in cancers~\cite{Ciriello,Miller}. Our
new algorithm can find the optimal solution for the problem, which
is better than solutions found by the heuristic algorithms in the
previous research~\cite{Ciriello,Miller}. The exclusivity is the
extreme case. In practical applications, a cancer cell may have
more than one mutation to perturb a common pathway. Hence, a
modified model is finding a set of mutations with minimum weight
sum such that each cancer cell has at least one and at most t (t=2
or 3) mutations in the solutions, which leads to the {\sc small
overlapped set cover} problem. Also, on application, some
mutations in cancer cells may not be detected because of errors.
Thus, it is not always ideal to find a solution mutations that
cover all cancer cells. A modified model is finding a set of
mutually exclusive mutations that cover at least $r$ percent
($90\%$ or $95\%$) of cancer cells, which leads to the {\sc
maximal set cover} problem. Our next research will design
efficient algorithms for above two new problems.


\end{document}